\documentclass[fleqn,usenatbib]{mnras}

\usepackage{newtxtext,newtxmath}

\usepackage[T1]{fontenc}

\DeclareRobustCommand{\VAN}[3]{#2}
\let\VANthebibliography\thebibliography
\def\thebibliography{\DeclareRobustCommand{\VAN}[3]{##3}\VANthebibliography}

\usepackage{graphicx}	
\usepackage{amsmath}	
\usepackage{xcolor}

\newcommand{\pcm}{\,cm$^{-2}$}	
\newcommand{\erg}{\,erg cm$^{-2}$ s$^{-1}$}	
\newcommand{\lum}{\,erg s$^{-1}$}	
\newcommand{\src}{\,4U 1626-67}	
\newcommand{\nus}{\,\emph{NuSTAR}}	

\usepackage[normalem]{ulem}

\title[New torque reversal of \src]{\src\ Returns to Spin-Down: Timing features toe the line}

\author[R. Sharma et al.]{
Rahul Sharma,$^{1}$\thanks{E-mail: rahul1607kumar@gmail.com; rsharma@rri.res.in (RS)}
Chetana Jain$^{2}$\thanks{E-mail: chetanajain11@gmail.com (CJ)} and Biswajit Paul$^{1}$
\\
$^{1}$Raman Research Institute, C V Raman Avenue, Sadashivanagar Bangalore 560080, INDIA\\
$^{2}$Hansraj College, University of Delhi, Delhi 110007, India\\}

\date{Accepted XXX. Received YYY; in original form ZZZ}

\pubyear{2023}

\begin{document}
\label{firstpage}
\pagerange{\pageref{firstpage}--\pageref{lastpage}}
\maketitle

\begin{abstract}

We present a comprehensive analysis of X-ray pulsar \src\ during its current spin-down (2SD) state, following a recent torque reversal. Since its discovery, this ultra-compact binary has experienced multiple torque states, transitioning from spin-up (1SU) during 1977--1990 to spin-down (1SD) during 1990--2008, and again spin-up (2SU) until 2023. From \nus\ observation of May 2023, we have investigated the timing and spectral properties of this pulsar during its 2SD phase, while also comparing them with previous spin-up-down states. For energies upto 8 keV, a distinct bi-horned pulse profile was observed during the spin-up phase, while several sub-structures emerged during spin-down. Beyond 8 keV, a broad asymmetric peak was consistently observed across all torque states. The pulse fraction during the 2SD phase was higher than that during 2SU phase. A prominent $\sim$46.8 mHz quasi-periodic oscillation has been exclusively detected during the spin-down phase. The broadband spectrum during the 2SD phase is described by empirical NPEX model, cyclotron absorption feature and its first harmonic. The spectrum during 2SU phase requires an additional blackbody component and asymmetry in the cyclotron absorption line. A significant flux drop by a factor of $\sim3$ in the 2SD was observed.

\end{abstract}




\section{Introduction}

\src\ is a $\sim$7.7 s accretion-powered X-ray pulsar in an ultra-compact ($\sim$42 min orbit) binary \citep{Middleditch81, Chakrabarty98}. Since its discovery in 1972 \citep{Giacconi72, Rappaport77}, this X-ray pulsar has shown several unique timing and spectral features associated with systematic changes in the inner accretion flow. 
The most prominent characteristic of \src\ is its steady spin-up and spin-down episodes of several years with torque reversals in between.

At first, \src\ was observed to be steadily spinning up until around the year 1990 (First spin-up: 1SU, hereafter), when it underwent a torque reversal to spin-down (First spin-down: 1SD, hereafter) \citep{Chakrabarty97}. After about 18 years, the pulsar witnessed another torque reversal to spin-up around the year 2008 (Second spin-up: 2SU, hereafter) \citep{Camero10, Jain10}. Recently, \src\ witnessed another torque reversal to spin-down \citep{Jenke23}. Historically, several timing and spectral characteristics of \src\ are known to be torque dependent, with similarities between 1SU and 2SU and differing from 1SD. 

Torque reversals in \src\ have been accompanied by significant changes in the source luminosity. During 1SU, the source luminosity was about 10$^{37}$ erg s$^{-1}$ \citep{White83}. Transition to the 1SD was accompanied by a decrease in the X-ray flux \citep{Chakrabarty97}. The source luminosity increased by a factor of about 2-3 during transition to the 2SU \citep{Jain10}. The latest torque reversal to spin-down (2SD) in \src\ is accompanied by a decrease in luminosity \citep{Jenke23}, which is similar to the first torque reversal that occurred in 1990. 

The X-ray pulse profile of \src\ is known to be strongly dependent on the torque state and energy \citep{Beri14}. The source has also displayed torque dependent mHz quasi-periodic oscillations (QPOs) in the power density spectrum (PDS) \citep{Kaur08, Jain10, Beri14}. During 1SU, 40 mHz weak QPOs were detected \citep{Shinoda90}. During 1SD, \citet{Owens97, Kommers98} and \citet{Chakrabarty98} reported strong QPOs at around 48 mHz. It was also observed that the QPO central frequency evolved slowly with time \citep{Kaur08}. QPOs were not observed during 2SU \citep{Jain10}.

The X-ray spectrum of \src\ is equally interesting. It is generally modelled with a blackbody and a hard power-law, along with the presence of Ne and O emission lines \citep{Angelini95}. The X-ray continuum is known to be correlated with the torque state. During 1SU, the spectrum had a blackbody temperature of about 0.6 keV and power-law photon index ($\Gamma$) of about 1.5 \citep{Pravdo79, Kii86}. The energy spectrum was relatively harder during 1SD, with $\Gamma \sim$0.5 and blackbody temperature was $\sim$0.3 keV \citep{Yi99}. The blackbody temperature after the second torque reversal (2SU phase) was similar to that reported during 1SU \citep{Jain10, Camero12}. 

The equivalent width and intensity of Ne and O emission lines are known to be variable with relatively higher intensity during the spin-up phase \citep{Camero12}. These emission lines also display strong pulse phase dependence indicating a warped structure in the accretion disk \citep{Beri15}. The presence of Fe K$_{\alpha}$ emission line is also strongly correlated with the luminosity of \src\ \citep{Koliopanos16} and has been detected only during the spin-up phase of \src.  

\citet{Orlandini98} discovered a narrow 37 keV cyclotron resonance scattering feature (CRSF) in the spectrum of \src\ during the 1SD. This was later confirmed by \citet{Coburn02} and \citet{Iwakiri12}, who also reported the presence of its harmonic from \textit{RXTE} and \textit{Suzaku} observations of 1SD, respectively. \citet{Iwakiri19} and \citet{DA17} reported asymmetry in the cyclotron line profile from data taken during the 2SU. 

Our objective for this letter is to investigate these similarities/dissimilarities in light of the recent torque reversal. In this work, we present the results from a recent \nus\ observation of \src\ made during 2SD. The results have been compared with those obtained during the previous torque states of the source.

\section{Observation and Data Reduction}

The Nuclear Spectroscopic Telescope ARray \citep[\nus;][]{Harrison13} mission comprises of two telescopes which focus X-rays between 3 and 78 keV onto two identical focal planes (usually called focal plane modules A and B, or FPMA and FPMB). \src\ was recently observed with \nus\ in May 2023. To investigate the temporal and spectral changes in the observed emission features w.r.t. results known until now, we have also used data from observations of \nus\ made in 2015 during the 2SU phase (see, Table \ref{tab:obs}). 

For this work, we have used the most recent \nus\ analysis software distributed with HEASOFT version 6.31.1 and the latest calibration files (version 20230124) for reduction and analysis of the \nus\ data. The calibrated and screened event files have been generated by using the task \textsc{nupipeline}. A circular region of radius 80 arcsec centred at the source position was used to extract the source events. Background events were extracted from a circular region of the same size away from the source. The task \textsc{nuproduct} was used to generate the light curves, spectra, and response files. Light curves were corrected to the Solar system barycentre using \textsc{barycorr}. The FPMA and FPMB light curves were background corrected and summed using \textsc{lcmath}. The spectra were grouped using \textsc{ftgrouppha} by following the \citet{Kaastra16} optimal binning scheme with a minimum of 25 counts per bin.

\begin{table}
    \centering
    \caption{The log of \nus\ observations.}
    \resizebox{0.9\columnwidth}{!}{
    \begin{tabular}{cccc}
    \hline
      Obs-ID & Date  & Exposure & Spin-phase \\
            & (yy-mm-dd) & (ks) & \\
    \hline  
       30101029002 & 2015-05-04 & 65 & 2SU \\
       90901318002 & 2023-05-02 & 27 & 2SD \\
    \hline     
    \end{tabular}}    
    \label{tab:obs}
\end{table}

\section{Results}

\subsection{Timing Analysis}

\begin{figure}
	\includegraphics[width=\columnwidth]{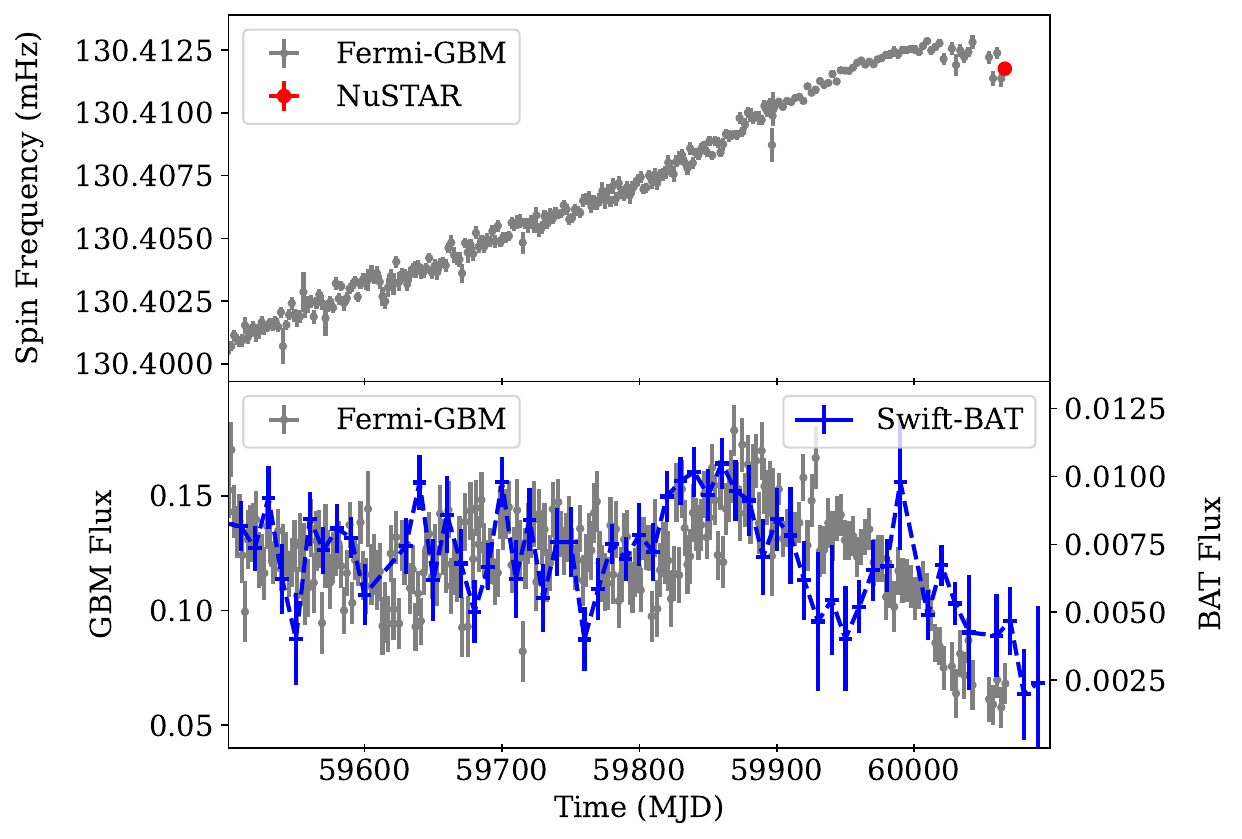}
    \caption{\emph{Top panel}: Spin frequency history of \src\ from Fermi-GBM. The spin period determined from \nus\ observation is shown with a filled red circle. \emph{Bottom panel}: The 12--25 keV pulsed flux (in units of keV \pcm\ s$^{-1}$) from Fermi-GBM and the 15--50 keV flux (in units of counts \pcm\ s$^{-1}$) from \emph{Swift}-BAT of \src. The \emph{Swift}-BAT data has been re-binned to 10 days for clarity.}
    \label{fig:gbm}
\end{figure}

We extracted 0.1 s lightcurve in the 3--78 keV energy range and searched for X-ray pulsations using the epoch folding method \citep{Leahy83}. We obtained a spin period of $P_{2015} = 7.672947 (1)$ s and $P_{2023} = 7.668020 (5)$ s for the 2015 and 2023 observations of \nus, respectively. The error in the spin period was estimated using the bootstrap method \citep{Lutovinov12, Boldin13}, by simulating 1000 light curves as described in \citet{Sharma23}. These results are consistent with Fermi/Gamma-ray Burst Monitor (GBM) pulsar program \citep{Malacaria20}. The top panel of Figure~\ref{fig:gbm} shows the spin frequency history of \src\ obtained from Fermi-GBM, along with the estimated value from \nus's recent observation. The trend in the spin frequency of \src\ suggests that torque reversal occurred around MJD 60020 (March 17, 2023) \citep{Jenke23}. The bottom panel of Figure~\ref{fig:gbm} shows 12--25 keV pulsed flux of \src\ from Fermi-GBM and 15--50 keV continuum flux from \emph{Swift}-BAT \citep{Krimm13}. Before MJD 59800, the temporal variation in the pulsed flux and total continuum flux overlapped. But beyond this, both the fluxes showed a considerable increase, with the pulsed flux lagging behind. After MJD 59900, the difference between the pulsed flux and continuum flux became more apparent. This might be correlated to alteration of emission geometry and hence the pulse profile as discussed later. Both the fluxes have reduced to about one-third of the peak flux seen around MJD 59870.

\begin{figure}
	\includegraphics[width=0.9\columnwidth]{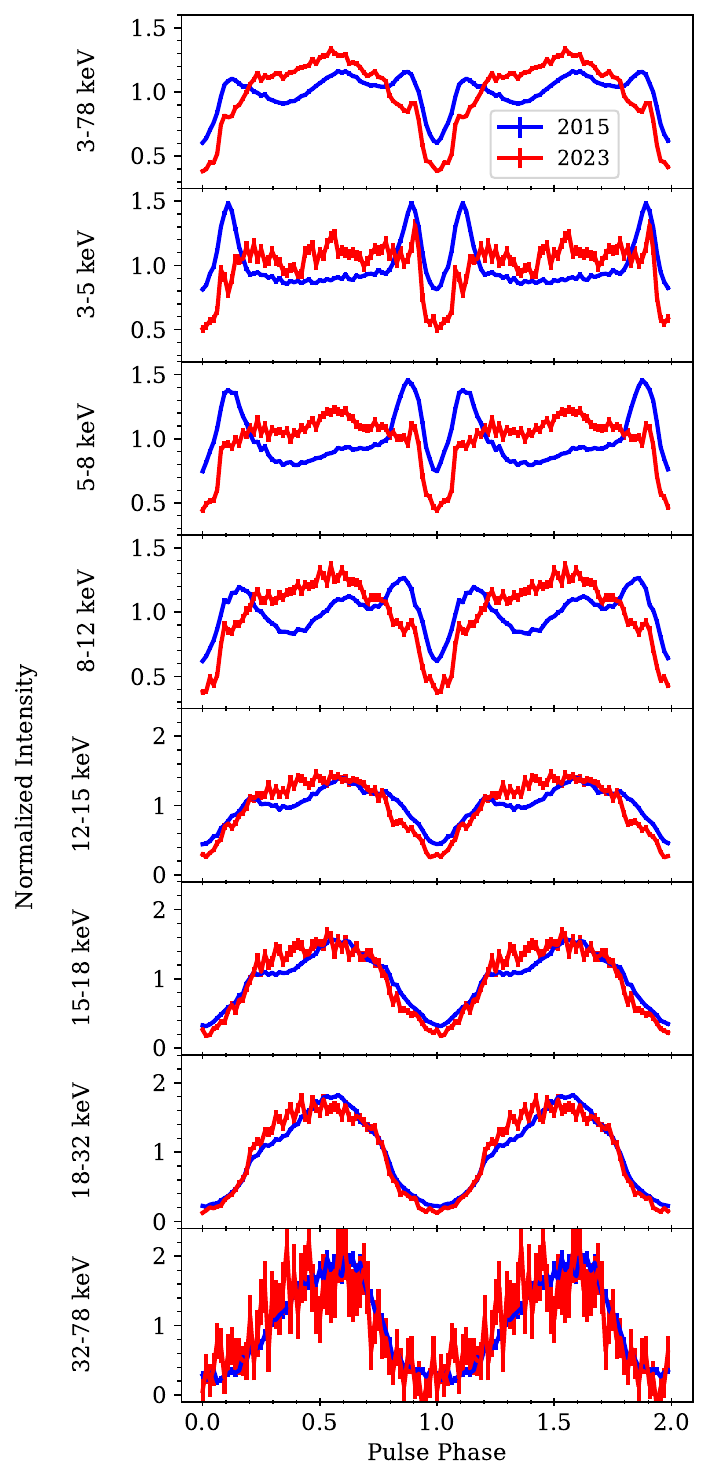}
    \caption{Energy-resolved pulse profile from two \nus\ observations of \src.}
    \label{fig:profile}
\end{figure}

The 3--78 keV light curves from both the \nus\ observations were folded with the corresponding pulse period to obtain pulse profiles. The energy-resolved  pulse profiles (3--5, 5--8, 8--12, 12--15, 15--18, 18--32 and 32--78 keV) are shown in Figure \ref{fig:profile}. In this figure, the blue and red colors are for the 2015 and 2023 observations, respectively. Clearly, the characteristics of the pulse profile are strongly dependent on energy as well as the torque state. Especially at low energies (up to 18 keV), the shape of the pulse profile has significantly changed since the 2SU. 

At energies up to 8 keV, the 2015 profile shows bi-horned profile with peaks separated by a narrow dip. Then up to about 15 keV, there appear two shallow dips at the pulse phase of 0.4 and 0.7. Thereafter, the profile tends to become broad single peaked. This is consistent with \citet{Beri14} and \citet{Iwakiri19}. The bi-horned profile below 8 keV is replaced by several sub-structures in the 2023 observation. Beyond 8 keV, the shape is largely a broad asymmetric peak and the width of main dip is seen to increase with energy. This is somewhat similar to profiles seen by \citet{Krauss07}. We have found that the 3--78 keV pulse fraction for 2023 profiles ($55.3 \pm 1.5$\%) is higher than that of 2015 profiles ($31.5 \pm 0.5$\%). 

\begin{figure}
	\includegraphics[width=0.9\columnwidth]{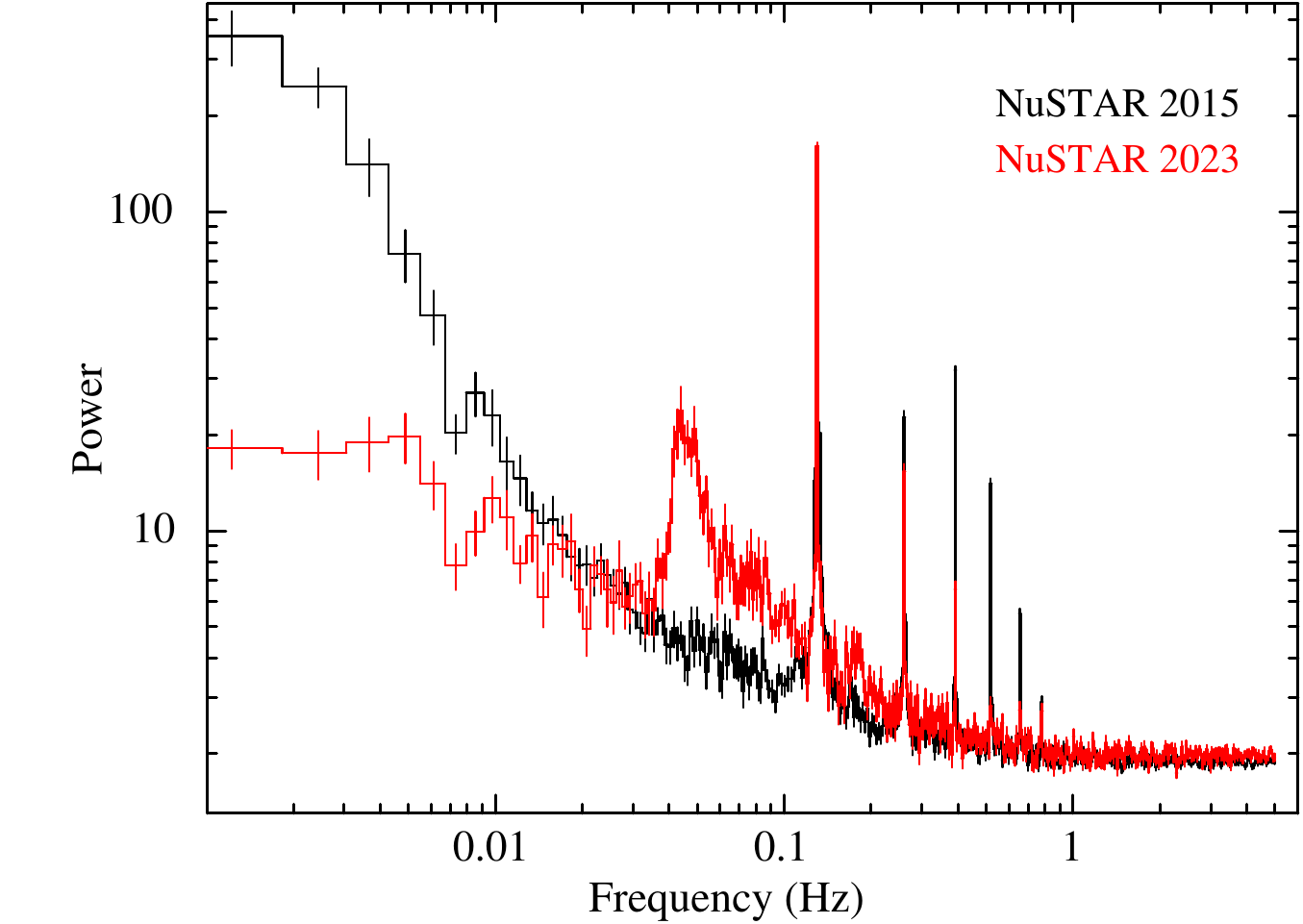}
    \caption{The 3--78 keV PDS of \src\ from \nus\ observations of 2015 and 2023.}
    \label{fig:pds}
\end{figure}

Figure~\ref{fig:pds} shows the 3--78 keV PDS of \src\ generated from 2015 and 2023 data by using the \textsc{powspec} tool of XRONOS sub-package of \textsc{ftools} \citep{Blackburn99}. Sharp peak at $\sim$0.13 Hz and its harmonics are visible in both observations. These correspond to spin period of $\sim$7.7 s. In addition, a broad peak corresponds to the quasi-periodic signal of $46.8 \pm 0.8$ mHz with r.m.s. amplitude of $16.1 \pm 1.5 \%$ and quality factor\footnote{Quality factor of a QPO is defined as the ratio between the QPO centroid frequency and FWHM \citep{Belloni02, vanderKlis04}.} of $\sim$5.9 is present only in the PDS of 2023 data. 

\subsection{Spectral Analysis}

We have used \textsc{xspec} \citep{Arnaud} version 12.13.0c for the spectral fitting. The 3--78 keV persistent spectra extracted from \nus-FPMA and \nus-FPMB observations were fitted simultaneously. We have added a constant to account for cross-calibration of the two instruments. The value of constant for \nus-FPMA was fixed to 1 and set free for FPMB. The photo-electric absorption cross-section of \citet{Verner} and abundance of \citet{Wilms} have been used throughout and $N_H$ was fixed to $9.6 \times 10^{20}$ cm$^{-2}$ \citep{HI4PI}. This Galactic survey value of $N_H$ is close to that observed with missions having a good soft energy coverage \citep[e.g.,][]{Camero12, Beri15, DA17}. All the spectral uncertainties and the upper limits reported in this paper are at a 90\% confidence level. We have assumed the source distance to be equal to 10 kpc, following \citet{Iwakiri19}.

We modelled the phase-averaged spectra of both the \nus\ observations using the Negative and Positive power-law with Exponential cutoff \citep[NPEX;][]{Mihara95, Makishima99} model. The 2015 spectra showed soft excess \citep{Schulz01, Iwakiri19}, which we modelled using thermal blackbody emission component \texttt{bbodyrad} of $kT_{\rm BB}=0.47$ keV. This soft excess was not detected in the 2023 spectra probably due to limited bandpass capability of \nus. We have estimated an upper limit on blackbody luminosity to be $7 \times 10^{34}$ \lum\ for $kT_{\rm BB} = 0.47$ keV. Considering $kT_{\rm BB}$ of 0.3 keV, as observed during 1SD \citep{Yi99, Beri15, Koliopanos16}, this upper limit changes to $8 \times 10^{35}$ \lum. Since the interstellar absorption is often poorly constrained in \nus, we checked whether the choice of hydrogen column density is correlated with the detection of blackbody component. For this purpose, the value $N_H$ was kept free during spectral fitting. We found an upper limit of $2.8 \times 10^{21}$ cm$^{-2}$ for the 2015 spectra and $6.5 \times 10^{21}$ cm$^{-2}$ for the 2023 spectra. We did not detect any difference in parameters of the blackbody component with/without freezing the column density. 

We have detected a broad Fe K$_\alpha$ emission line at $6.75 \pm 0.05$ keV during 2SU observation of 2015, with an equivalent width of $\sim 26$ eV, consistent with previous known results \citep{Koliopanos16, Iwakiri19}. In 2023 spectra, the Fe emission line was tentatively detected with a similar equivalent width of $\sim 26$ eV. The Gaussian width was fixed at 0.15 keV due to unconstrained values. The addition of this Gaussian component improved the spectral fit for the 2023 observation, with a $\Delta \chi^2$ of 13 for two additional degrees of freedom (dof) with a false alarm probability of 0.11\%, which corresponds to the statistical significance of 3$\sigma$. The Fe emission feature was not detected during the previous spin-down phase of \src.  

Since the source is known to show asymmetric CRSF feature around 37 keV \citep{DA17, Iwakiri19}, we tried two models - \texttt{gabs} and \texttt{cyclabs} in \textsc{xspec} to investigate the shape of the absorption line. The \texttt{gabs} model has a Gaussian optical-depth profile, whereas, the \texttt{cyclabs} model has a pseudo-Lorentzian optical-depth profile \citep{Mihara90, Makishima90}. Although, both the models gave statistically similar fit, the Lorentzian line profile along with its width and depth described the spectrum marginally better than \texttt{gabs} model. Therefore, we used \texttt{cyclabs} to model the fundamental CRSF feature. 

In addition to the fundamental CRSF, we observed additional absorption feature at energies above 60 keV, which might correspond to  CRSF harmonic \citep{Coburn02, DA17}. Since the spectrum has relatively poor statistics above 60 keV, therefore, we modelled this possible harmonic feature using the \texttt{gabs} model. Though the 2023 spectra showed additional residuals around 66 keV, but owing to few bins and insufficient statistics, we fixed the line energy and width of this feature to that determined in the 2015 spectra. This gave an upper limit of 152 on the line depth which corresponds to an optical depth of $\sim 7$. The fit statistics improved by $\Delta \chi^2 =39$ for 3 additional dof. Thus, the 2023 spectra is best described by \texttt{tbabs*(NPEX+Gaussian)*cyclabs*gabs} model.

The 2015 spectra showed an additional dip in residuals near 40 keV \citep{Iwakiri19}, which was not seen in the 2023 spectra. We added another Gaussian absorption component to account for an asymmetry of the cyclotron line. This improved the fit statistics by $\Delta \chi^2 = 27$ for 3 additional dof. Thus, the 2015 spectral model is best described by \texttt{tbabs*(bbodyrad+NPEX+Gaussian)*cyclabs*gabs*gabs}. 

To test for the significance of all the Gaussian absorption components, we used \texttt{simftest} script of \textsc{xspec}, which is based on Monte Carlo simulation of data sets having same counting statistics as the original data. We simulated 10,000 data sets which were fit with best-fit spectral models with the component to be tested and without it. The 66 keV \texttt{gabs} component has a false alarm probability of $< 10^{-4}$ for both 2015 and 2023 observations, corresponding to the statistical significance of $>3 \sigma$. Similar significance was found for the 40 keV absorption feature in the case of 2015 spectra. Contrary to \citet{Iwakiri19}, we did not require additional broad Gaussian at 20 keV to model broad hump feature. 

\begin{figure}
	\includegraphics[width=\columnwidth]{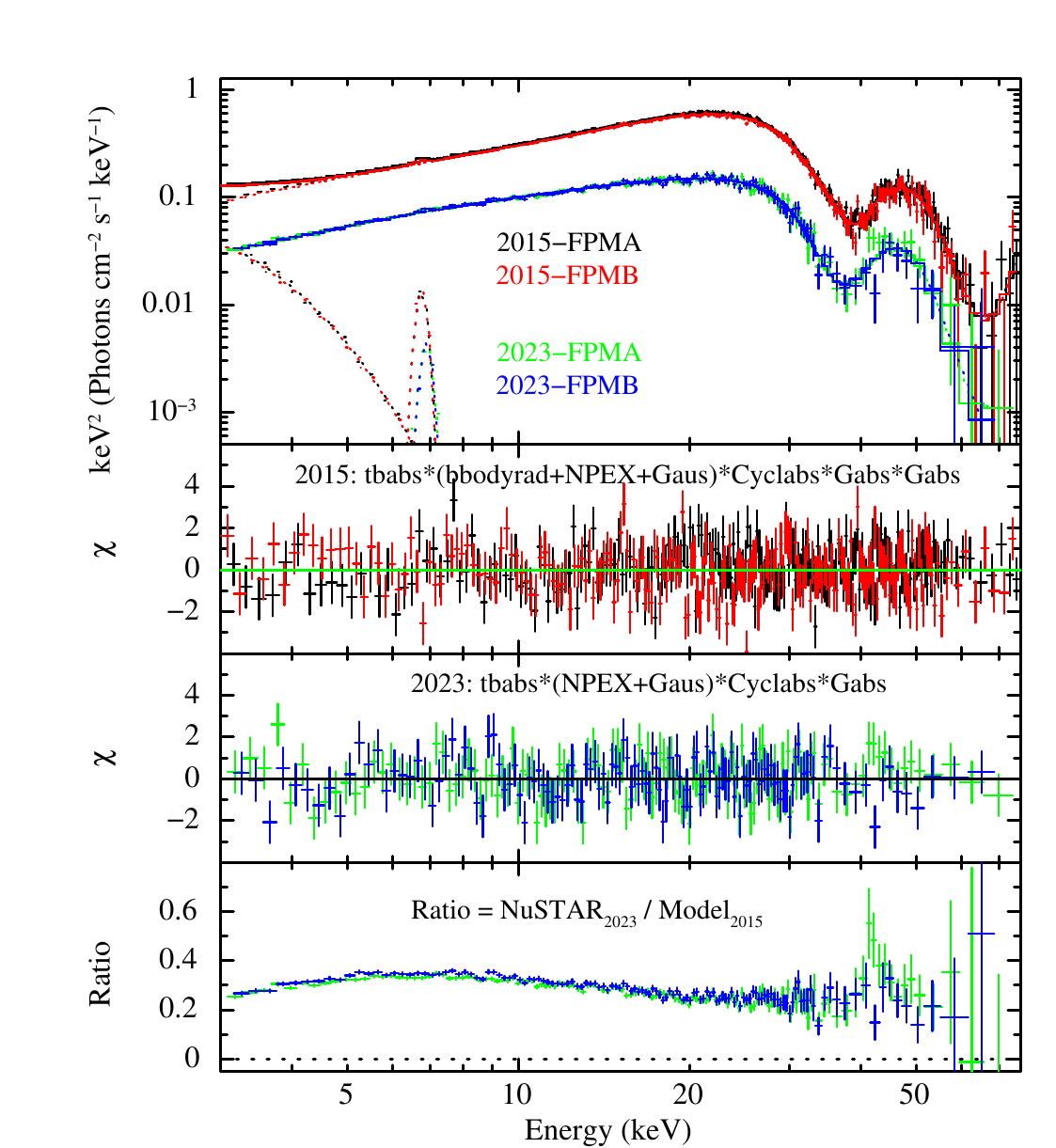}
    \caption{The unfolded spectrum of two \nus\ observations are shown in the top panel for comparison and fitted with their respective best-fit model as given in Table \ref{tab:spec}. The second and third panels show the residue with the best-fit model for 2015 and 2023 \nus\ observations, respectively. The bottom panel show the ratio of the \nus\ 2023 data with the best-fit spectral model of 2015, suggesting the flux decrease by a factor of 3 during the 2023 observation. }
    \label{fig:spec}
\end{figure}

The best-fit model with their respective residuals is shown in Figure~\ref{fig:spec}, and parameters of the best-fit model are given in Table~\ref{tab:spec}. Compared to the 2015 \nus\ observation, the 3--78 keV unabsorbed flux during the recent observation has reduced to $\sim$27$\%$ of its value during the observation in 2SU. The bottom panel of Figure~\ref{fig:spec} shows the ratio of the \nus\ 2023 data with the best-fit spectral model of 2015 data. The ratio is nearly around 0.3 and shows a flat trend at higher energy. The spectral variability across the spectra of the two torque states is not very significant. Only the flux has decreased by a factor of $\sim 3$ with an absence of soft thermal component.

\renewcommand{\arraystretch}{1.3} 
\begin{table}
	\centering
	\caption{Best-fit spectral parameters of \src\ during 2015 and 2023 observations. All errors and upper limits reported in this table are at $90\%$ confidence level ($\Delta \chi^2=2.7$).} 
	\label{tab:spec}
	\resizebox{\columnwidth}{!}{
	\begin{tabular}{lccr} 
		\hline
		Component & Parameters & \nus-2015$^a$ & \nus-2023$^b$ \\
		\hline
		TBabs & $N_{\rm H}$ ($10^{20}$ \pcm) & \multicolumn{2}{c}{$9.6^{\rm fixed}$} \\
	     
		BBodyrad & $kT_{\rm BB}$ (keV) & $0.475 \pm 0.03$ & - \\
		         & Norm & $251^{+134}_{-82}$ & - \\
              & $L_{\rm BB}$ ($10^{36}$ \lum) & $1.64_{-0.21}^{+0.30}$ & \\
		         
		NPEX & $\Gamma$ & $0.67_{-0.03}^{+0.09}$ & $0.53^{+0.14}_{-0.08}$ \\
		        & $f$ ($10^{-4}$) & $7.26_{-3.2}^{+1.9}$ & $4.4^{+4.6}_{-3.1}$\\
		        & $E_{\rm cut}$ (keV) & $10.5^{+4}_{-1}$ & $9.95^{+4.8}_{-2.0}$\\
		        & Norm ($10^{-2}$) & $2.92 \pm 0.15$ & $0.82^{+0.06}_{-0.03}$ \\

        Gaussian &  $E_{\rm Fe~K}$ (keV) & $6.75 \pm 0.05$ & $6.89^{+0.30}_{-0.19}$\\
            & $\sigma_{\rm Fe~K}$ (keV) & $0.14^{+0.07}_{-0.08}$ & $0.15^{\rm fixed}$\\
            & Eqw (eV) & $26 \pm 6$ & $26_{-11}^{+20}$ \\
            & Norm ($10^{-4}$) & $1.23^{+0.28}_{-0.26}$ & $0.40 \pm 0.18$ \\
          
        Cyclabs & $E_{\rm cyc}$ (keV) & $35.9^{+0.3}_{-0.7}$ & $35.9 \pm 0.4$ \\
                  & $\sigma_{\rm cyc}$ (keV) & $9.39^{+2.76}_{-0.66}$ & $6.86^{+0.95}_{-1.0}$ \\
                  & Depth & $2.52^{+0.21}_{-0.25}$ & $2.5 \pm 0.3$ \\

        Gabs &  $E_{1}$ (keV) & $39.23^{+0.62}_{+1.05}$ & - \\
            & $\sigma_{1}$ (keV) & $2.2^{+1.2}_{-0.7}$ & - \\
            & Norm$_1$ & $1.97^{+1.91}_{-0.95}$ & - \\

        Gabs &  $E_{2}$ (keV) & $65.9^{+2.2}_{-1.2}$ & $65.9^{\rm fixed}$ \\
            & $\sigma_{2}$ (keV) & $9^{+3}_{-1}$ & $9^{\rm fixed}$ \\
            & Norm$_2$ & $89.4^{+23}_{-18}$ & $101^{+51}_{-38}$ \\         
		        
		        & $C_{\rm FPMB}$ & $0.965 \pm 0.002$ & $1.0005 \pm 0.007$ \\

        Flux$^c$   & $F_{3-78~{\rm keV}}$ & $1.35 \times 10^{-9}$ & $3.7 \times 10^{-10}$ \\
                   & $F_{0.1-100~{\rm keV}}$ & $1.65 \times 10^{-9}$ & $4.15 \times 10^{-10}$ \\

		\hline        
		        & $\chi^2$/dof & 460/383 & 240.5/258 \\      
		\hline
		\multicolumn{4}{l}{$^a$ Model of 2015: \texttt{tbabs*(bbodyrad+NPEX+gaus)*Cyclabs*Gabs*Gabs}.}\\
		\multicolumn{4}{l}{$^b$ Model of 2023: \texttt{tbabs*(NPEX+gaus)*Cyclabs*Gabs}.}\\
        \multicolumn{4}{l}{$^c$ Unbsorbed flux in units of \erg.}\\
	\end{tabular}}
\end{table}

\section{Discussion and Conclusions}

This work reports the timing and spectral properties of \src\ after its recent torque reversal to the spin-down phase and comparison with the previous spin-up phase using \nus\ observations of 2015 and 2023. Coherent pulsations at $\sim 7.67$ s were detected in both phases. The analysis reveals that the pulse profiles exhibit energy and torque-state dependence during both observations. Notably, the double-peaked structure observed in the pulse profile disappears during the spin-down phase. The pulse profile shape is in accordance with the previously known torque-state dependence patterns \citep{Beri14}. 
The PDS (Fig. \ref{fig:pds}) of the two observations shows starking dissimilarities. The 2SU observation of 2015 had higher power at frequencies below 0.01 Hz. The 2023 observation made during spin-down phase is characterised by the presence of a broad $46.8 \pm 0.8$ mHz QPO with a quality factor of $\sim 5.9$ and r.m.s. power of $\sim 16.1 \%$. This feature occurs exclusively during the spin-down episodes and is absent during the spin-up \citep{Kaur08, Jain10, Beri14}. The presence of strong QPO only in the spin-down era highlights the existence of two distinct accretion modes onto the pulsar, which correspondingly produce different pulse profile shapes \citep{Krauss07, Jain10, Beri14}. 

Contrary to the generally accepted spectral model for \src, a blackbody component is either not present in the current torque state or is not detectable with \nus\ owing to low energy threshold of 3 keV. 

We have determined a cyclotron line feature at $35.9 \pm 0.4$ keV, which implies a surface magnetic field strength of $\sim 3.1 \times 10^{12}$ G. We have also detected absorption at $\sim 66$ keV in both the observations with more than $3\sigma$ significance. The 66 keV feature can be due to harmonic of CRSF. But it is not exactly the double of CRSF observed at 35.9 keV. CRSFs with non-harmonic spacing have been reported in several other sources \citep[e.g.,][]{Heindl1999, Pottschmidt2005, Furst2018, Sharma22}. These have been attributed to various reasons such as, relativistic effects, scattering in different magnetic field regions, optical depth effects in the mound of accreting matter and lines produced at opposite magnetic poles. Historically, this harmonic feature has been reported by \citet{DA17}, but \citet{Iwakiri19} found this feature to be insignificant using the same 2015 \nus\ observations. The difference may lie in the continuum used. \citet{Iwakiri19} used modified NPEX model (NPEX + Gaussian for broad hump at 20 keV). We found that NPEX succeeded in modelling the continuum of both \nus\ observations. 
The 2015 spectra also showed an additional 40 keV absorption feature on top of CRSF \citep{Iwakiri19}. 

In the standard model of accretion onto magnetised neutron stars, accretion torque is an important marker which reflects the magnetic coupling between the neutron star and the accretion disc around it. Accretion torque is related to the mass accretion rate onto the neutron star. And changes in the accretion torque are expected to be related to the source luminosity and spin transition \citep{Ghosh79}. The timescale for torque reversals in \src\ can possibly be due to the presence of long-term cycles in the supply of matter from the companion star.

Torque reversal in \src\ suggests that the pulsar is spinning near its equilibrium period ($P_{\rm eq}$), where the spin frequency matches the Kepler frequency at the magnetosphere. A neutron star with a spin period shorter than $P_{\rm eq}$ cannot accrete easily and may experience a strong spin-down torque -- the propeller effect \citep{Illarionov75}. Using eqn (9) of \citet{Bildsten97}, we have estimated the average mass accretion rate of $\dot{M}_{\rm avg} \sim 10.6 \times 10^{-10} ~M_\odot$ yr$^{-1}$ using the magnetic field of $3.1 \times 10^{12}$ G. A mass accretion rate above $\dot{M}_{\rm avg}$ will result in spin-up torque while a lower accretion rate will cause spin-down torque. Using 0.1--100 keV un-absorbed flux, we have estimated an accretion rate of $\sim 12 \times 10^{-10} ~M_\odot$ yr$^{-1}$ for 2015 observation and $\sim 3 \times 10^{-10} ~M_\odot$ yr$^{-1}$ for 2023 observation. 

The characteristic torque $N_0 \equiv \dot{M} \sqrt{G M_X R_{co}}$ \citep{Bildsten97} is often determined by the mass accretion rate and specific angular momentum of matter at the corotation radius. This torque gives an estimate of the spin evolution rate ($\dot{\nu} \sim N/2 \pi I$ Hz s$^{-1}$). Assuming moment of inertia, $I\sim10^{45}$ g cm$^2$, $M_X = 1.4 M_\odot$, and $R_{co} = 6.5 \times 10^8$ cm \citep{Chakrabarty97}, we have estimated time scale of $\nu/\dot{\nu} \sim$2800 yr for the curent spin-down phase and $\sim$700 yr for the previous spin-up phase. For comparison, \src\ was spinning up steadily during 1977-1991 and made a transition to a steady spin-down at nearly the same rate of $\nu/\dot{\nu} \sim$5000 yr by 1991 \citep{Bildsten97}. But, despite this torque reversal, there was no evidence of a large change in the bolometric flux from the source \citep{Bildsten97}. Over a period of $\sim 300$ days, covering the pre- and post-current torque reversal, the hard X-ray flux has already dropped by a factor of $\sim 3$.

More observations of \src\ with \nus\ and several other X-ray missions are encouraged during the current phase in order to investigate the complex interaction between the accretion flow and the neutron star magnetosphere. 

\section*{Acknowledgements}

This research has made use of data obtained from the High Energy Astrophysics Science Archive Research Center (HEASARC), provided by NASA’s Goddard Space Flight Center. This research has made use of the \nus\ Data Analysis Software (NUSTARDAS) jointly developed by the ASI Science Data Center (ASDC, Italy) and the California Institute of Technology (USA). This research has also used data from Fermi-GBM pulsar monitor and \emph{Swift}-BAT transient monitor. We also thank the anonymous referee for insightful comments and suggestions.

\section*{Data Availability}

Data used in this work can be accessed through the High Energy Astrophysics Science Archive Research Center (HEASARC) archive at \url{https://heasarc.gsfc.nasa.gov/cgi-bin/W3Browse/w3browse.pl}. 



\bibliographystyle{mnras}
\bibliography{sample} 





\bsp	
\label{lastpage}
\end{document}